\documentstyle[12pt]{article}
\newcommand{\be}{\begin{equation}}
\newcommand{\ee}{\end{equation}}
\newcommand{\ba}{\begin{eqnarray}}
\newcommand{\ea}{\end{eqnarray}}

\begin{document}
\begin{center}
 {\bf\Large{
 Reparametrization invariance and Hamilton-Jacobi formalism
    }}
\end{center}
\begin{center} {\bf Dumitru
Baleanu\footnote[1]{E-Mails:~~dumitru@cankaya.edu.tr,~~baleanu@venus.nipne.ro}}
\end{center}
\begin{center}
Department of Mathematics and Computer Sciences, Faculty of Arts
and Sciences, Cankaya University-06530, Ankara , Turkey
\end{center}
\begin{center}
and
\end{center}
\begin{center}
Institute of Space Sciences, P.O. BOX, MG-23, R 76900,
Magurele-Bucharest, Romania
\end{center}

\begin{abstract}

 Systems invariant under the reparametrization of time were treated as  constrained
 systems within Hamilton-Jacobi formalism.
After imposing the integrability conditions the time-dependent
Schr\"odinger equation was obtained. Three examples are
investigated in details.
 \end{abstract}

\section{Introduction}

  Reparametrization theories of gravity such as general relativity
and string theory are invariant under reparametrization of time
\cite{dirac,henneaux, gitman, hartle}.  A transformation from a
reparametrization-invariant system to an ordinary gauge system was
applied for deparametrizing cosmological models \cite{simeone1}.
1By adding a surface term to the action functional the gauge
invariance of the systems whose Hamilton-Jacobi equation is
separable was improved \cite{simeone2}. Reparametrization
invariance was treated as a gauge symmetry in \cite{gitman} and
 a time-dependent Schr\"odinger equation for systems
invariant under the reparametrization of time was developed
\cite{Pashnev1,Pashnev2}. To reach this goal an additional
invariance action was introduced without changing the equation of
motions but modifying the set of constraints
 \cite{Pashnev1,Pashnev2}.
 After the second class constraints were eliminated one of first class constraints
 becomes time-depending Schr\"odinger equation.

  An intriguing question arises: could we construct a time-depending
  Schr\"odinger equation without involving any gauge invariance
  transformations for the extended action proposed in \cite{Pashnev1,Pashnev2}?

 The answer of this question is to use the Hamilton-Jacobi (HJ) formalism for constrained systems
\cite{guler1992},  based on Caratheodory's equivalent Lagrangians
method \cite{cara}. This formalism does not differentiate between
the first and second class constraints, we do not need any gauge
fixing terms and the action provided by (HJ) is useful for the
path integral quantization method of the constrained systems. In
addition, it was proved that the integrability conditions of (HJ)
formalism and Dirac's consistency conditions are equivalent
\cite{pimentel} and the equivalence of the chain method
\cite{mitra} and (HJ) formalism was investigated \cite{baleanu0}.

 The main aim of this paper is to obtain, using (HJ) formalism,
 the time-dependent Schr\"odinger equation for  theories invariant
 under the time
 reparametrization.

 The plan of the paper is as follows:\\
 In Sec. 1 the (HJ) formalism is briefly presented.
 In Sec. 2 the reparametrization invariance is investigated with
  (HJ) approach. Sec.3 is dedicated to our conclusions.

\section{Hamilton-Jacobi formalism}

(HJ) formalism based on Caratheodory's equivalent Lagrangians
method \cite{cara} is an alternative method \cite{guler1992} of
quantization of constrained systems. The basic idea of this new
approach is to consider the constraints as new "Hamiltonians" and
to involve all of them in the process of finding the action. Let
us consider that a given degenerate Lagrangian L admits the
following primary "Hamiltonians"
\begin{equation}
H'_{\alpha}=p_{\alpha}+H_{\alpha}(t_{\beta},q_{a},p_{a}).
\end{equation}
The canonical Hamiltonian  $H_{0}$ is defined  as
\begin{equation}
H_{0}=-L(t,q_{i},\dot{q_{\nu}},\dot{q_{a}}=w_{a})+p_{a}w_{a}+\dot{q_{\mu}}p_{\mu};\quad
\nu=0,n-r+1,....,n.
\end{equation}
The equations of motion are obtained as total differential
equations in many variables as follows:
\begin{equation}\label{equat}
dq_{a}=\frac{\partial H'_{\alpha}}{\partial
p_{a}}dt_{\alpha},\quad dp_{a}=-\frac{\partial
H'_{\alpha}}{\partial q_{a}}dt_{\alpha},\quad
dp_{\mu}=-\frac{\partial H'_{\alpha}}{\partial
t_{\mu}}dt_{\alpha},\quad \mu=1,...,r.
\end{equation}
and  the (HJ) function is given by
\begin{equation}
dz=(-H_{\alpha}+p_{\alpha}\frac{\partial H'_{\alpha}}{\partial
p_{a}})dt_{\alpha}
\end{equation}
 The set of equations (\ref{equat}) is integrable if
and only if \cite{guler1992}
\begin{equation}\label{inte}
[H'_{\alpha},H'_{\beta}]=0 \quad \forall\alpha ,\beta.
\end{equation}

The method is straightforward for constrained systems having
finite degree of freedom \cite{baleanu} but it becomes in some
cases quite difficult to be used  for field theories. The main
difficulty comes from the fact that some surface terms may play an
important role in closing the algebra of "Hamiltonians" but some
of them have no physical meaning from the (HJ) point of view.
Another problem is the treatment of the second-class constraints
systems within (HJ) formalism \cite{rabei1992}. In this particular
case the "Hamiltonians" are not in involution and it is not a
unique way to solve this problem  \cite{baleanu1, baleanu2,
baleanu3}.

\section{Reparametrization invariance}

    First of all we mention that if we apply naively the (HJ)
formalism to reparametrization invariance theories no
time-dependent Schr\"odinger equation appears simply because of
the equivalence of  (HJ) and Dirac's formalisms.

In order to obtain a time-dependent Schr\"odinger equation the
initial Lagrangian is extended with a term involving the lapse
function N and in the last example with its corresponding
superpartners. In this manner a second class-constrained system is
obtained. This system is investigated within (HJ) formalism and
after imposing the integrability conditions we obtained the
time-dependent Schr\"odinger equation on the surface of
constraints.

\subsection{Examples}

\subsubsection{Non-relativistic particle dynamics }

 The Lagrangian for a non-relativistic particle moving in the
 three dimensional space is given by
\begin{equation}\label{lkl}
L=\frac{1}{2}m\dot{x_i}^{2}(t)-V(x_{i}),
\end{equation}
where $x_{i}, i=1,2,3$ are dynamical variables, t denotes the
ordinary physical time parameter, m is the mass of the particle
and $V(x_{i})$\quad is the potential.

Treating the time as a dynamical quantity the Lagrangian
(\ref{lkl}) takes the form
\begin{equation}
L^{'}=\frac{1}{2N(\tau)}m\dot{x_{i}}^{2}(t)-N(\tau)V(x_{i}(\tau))
\end{equation}
where $\dot{x_{i}}=\frac{dx_{i}}{d\tau}$\quad and $N(\tau)$\quad
is the lapse function relating the physical time to the arbitrary
parameter  by $dt=N(\tau)d\tau$.

In order to obtain a time-dependent Schr\"odinger equation we
consider the extended Lagrangian

\be\label{asa1} L^{''} =
\frac{1}{2N(\tau)}m\dot{x_{i}}^{2}(t)-N(\tau)V(x_{i}(\tau))
+\lambda(-N+{dt\over d\tau})
 \ee
From (\ref{asa1}) we obtain \be\label{asa2} p_i=m{{\dot x_i}\over
N}, p_\lambda=0,p_N=0, p_t=\lambda. \ee

Using (\ref{asa1}) and (\ref{asa2}) the expression for canonical
Hamiltonian becomes

\be H_c= N({{p_i}^2\over 2m}-V(x_i)+\lambda) +{\dot
t}(p_t-\lambda) \ee

Therefore in (HJ) formalism we have the following "Hamiltonians":

\be\label{unspe} H_0^{'}=p_0+ N({{p_i}^2\over 2m}-V(x_i)+\lambda),
H_1^{'}=p_\lambda,H_2^{'}=p_N,H_3^{'}=p_t-\lambda
 \ee

We observe from (\ref{unspe}) that the system of second class
constraints in Dirac's classification and $H_c$ is a combination
of $H_0={{p_i}^2\over 2m}-V(x_i)$, $\lambda$ and $H_3^{'}$. To
make the system integrable we  impose the integrability
conditions. Making the variation of $H_2^{'}$ zero we obtained
another "Hamiltonian" \be\label{doispe}
H_4^{'}=\lambda+{{p_i}^2\over 2m}-V(x_i). \ee On the surface of
constraints it is easy to realize that $p_t=\lambda$ and
(\ref{doispe}) represents the time-dependent Schr\"odinger
equation.

\subsection{Relativistic point particle}

The well known Lagrangian for a free
relativistic particle is given by
\begin{equation}\label{lala}
L=-m\sqrt{1-\dot{x}_{i}^{2}}.
\end{equation}
Here m, t and $x_{i}$\quad are, the mass, proper time and the
position of the particle. If we make the reparametrization
transformation

\begin{equation}
dx^0=N(\tau)d\tau,
\end{equation}

from (\ref{lala}) it follows that

\begin{equation}\label{lalaa}
L^{'}=-m\sqrt{N^{2}(\tau)-\dot{x}_{i}^{2}}.
\end{equation}

 To obtain a time-dependent Schr\"odinger equation we consider the
 extension of (\ref{lalaa}) as

 \be\label{alal}
 L^{''}= -m\sqrt{N^{2}(\tau)-\dot{x}_{i}^{2}}+ \lambda ({\dot x^0-N})
 \ee

From (\ref{alal}) we obtain

\begin{equation}\label{llaa}
p_{i}=m\frac{\dot{x}_{i}}{\sqrt{N^2-\dot{x}_{i}^2}}, p_\lambda=0,
p_N=0, p_{x^0}=\lambda
\end{equation}

Taking into account (\ref{alal}) and (\ref{llaa}),
 the corresponding canonical Hamiltonian is
\begin{equation}\label{can}
H_{c}=N(\sqrt{p_{i}^{2}+m^{2}}+\lambda)+ \dot
x^0(p_{x^0}-\lambda).
\end{equation}

In this case the "Hamiltonians"  of (HJ) formalism  are given by
\be\label{ddd} H_0^{'}=p_0+ N(\sqrt{p_{i}^{2}+m^{2}}+\lambda),
H_1^{'}=p_\lambda,H_2^{'}=p_N,H_3^{'}=p_{x^0}-\lambda \ee
 Since the set  from (\ref{can}) and  (\ref{ddd}) is not integrable in (HJ)
 formalism, we used the same procedure as in the previous example  and on the surface of constrains, the time-dependent Schr\"odinger
 equation is obtained as

 \be
 H_4^{'}=p_{x^0} + \sqrt{p_{i}^{2}+m^{2}},
\ee which  represents our desired result.

\subsection{N=2, D=1 supersymmetry}

In \cite{Pashnev2} the N=2 supersymmetric quantum mechanics was
coupled to world-line supergravity and the local supersymmetric
action was obtained. The action to start with is \be
S=S_{1(n=2)}+S_{2(n=2)}, \ee where

\be\label{s1} S_{1(n=2)}=\int\{{(Dx)^2\over 2N}-i{\bar\chi}D\chi
-2N({\partial g\over\partial x})^2- 2N{\bar\chi}\chi
{\partial^2g\over\partial x^2}+{\partial g\over\partial
x}({\bar\psi}\chi-\psi{\bar\chi})\}, \ee

\be\label{s2} S_{2(n=2)}=\int\{\lambda(-N+{\dot t}+{\bar\psi\over
2}{\bar\eta} -{\psi\over 2}{\eta})-\lambda_1( i{\dot\eta}+{V\over
2}\eta +{\bar\psi\over
2})+\lambda_2(-i{\dot{\bar\eta}}+{\bar\psi\over 2}+ {V\over
2}\bar\eta) \} \ee

Here the covariant derivatives are given by $Dx={\dot x}-{i\over
2}(\psi{\bar\chi}+{\bar\psi}\chi)$ and $D\chi={\dot\chi}+{i\over
2}V\chi$ respectively. $\lambda_1$ and $\lambda_2$ are the
superpartners of $\lambda$ and $\psi$,$\bar\psi$ and V are
superpartners of N.For more details concerning the above action
see \cite{Pashnev2} and the references therein. We observed that
in addition to the primary "Hamiltonians" \be\label{first}
H_1^{'}=p_N, H_2^{'}=p_{\psi},H_3^{'}=p_{\bar\psi}, H_4^{'}=p_V,
\ee from (\ref{s1}) and (\ref{s2}) we obtained a new set of
"Hamiltonians" having the following forms \be\label{second}
H_5^{'}=p_{\lambda}, H_6^{'}=p_{t}-\lambda,H_7^{'}=p_{\lambda_1},
H_8^{'}=p_{\lambda_2},
H_9^{'}=p_{\eta}+i\lambda_1,H_{10}^{'}=p_{\bar\eta}+i\lambda_2.
\ee By inspection we observed that the set of "Hamiltonians" from
(\ref{second}) is not in involution, so the system of total
differential equations corresponding to (\ref{first}) and
(\ref{second}) is not integrable within (HJ) formalism. The
canonical Hamiltonian becomes \be H_c=N(\lambda+H_0)-{\psi\over
2}(S_{\bar\eta}+{\bar S})+{\bar\psi}(-S_{\eta}+S) +{V\over
2}(F_\eta +F), \ee where \ba &H_0&={p^2\over 2}+2({\partial
g\over\partial x})^2+ 2({\partial^2 g\over\partial x^2}){\bar\chi
\chi},S_{\eta}=({\bar\eta p_t-p_\eta}),
S_{\bar\eta}=(p_{\bar\eta}-\eta p_t),\cr
 &F_\eta&=(\eta p_\eta
-{\bar\eta}p_{\bar\eta}), F={\bar\chi}\chi, S=(ip+2{\partial
g\over\partial x})\chi, {\bar S}=(-ip+2{\partial g\over\partial
x}){\bar\chi} \ea

 The "Hamiltonians" from (\ref{first}) and (\ref{second}) together
 with
\be\label{third}
 H_0^{'}=p_0+H_c
\ee
  form the total set  "Hamiltonians" for our investigated problem.
  The next step is to investigate the corresponding integrability conditions.
  Firstly we impose the variations of "Hamiltonians" given by
  (\ref{first}) to be zero.
  As a result we obtained  a new set of "Hamiltonians" as given below

\be\label{newham} H_{0}^{''}=\lambda+H_0, Q_\eta=-S_\eta +S,
Q_{\bar\eta}= S_{\bar\eta} +{\bar S}, {\tilde F}=F_\eta +F \ee
  On the surface of constraints of (\ref{second}) the set of
  "Hamiltonians"from (\ref{newham}) is in involution and
   we obtained  $H_0^{''}=p_t+ H_0$ which represents the time-dependent Schr\"odinger
   equation.

\section{Conclusions}

In this paper we proved that a time-dependent Schr\"odinger
equation was obtained for a given reparametrization theory by
using an extended Lagrangian involving the lapse function N and
its superpartners. Taking into account that  (HJ)  and Dirac's
formalisms are equivalent we observed that the reparametrization
invariance theories are not integrable inside of the (HJ)
formalism. This result is due to the fact that the integrability
conditions (\ref{inte}) are not satisfied. On the surface of
constraints we obtained the time-dependent Schr\"odinger equation
without involving any symmetry transformation of the extended
action.

 \section {Acknowledgments}
 This work is partially supported by the Scientific and
Technical Research Council of Turkey.

\end{document}